\documentclass[runningheads]{llncs}
\pdfoutput=1 
\usepackage{wrapfig}
\usepackage{palatino}
\RequirePackage[colorlinks=false]{hyperref}
\usepackage[shortlabels]{enumitem}
\usepackage{xcolor}

\newcommand{\coop}[1]{\langle{#1}\rangle}
\usepackage{enumitem}
\definecolor {processblue}{cmyk}{0.96,0,0,0}
\RequirePackage{amsmath,amssymb,ifthen,url,graphicx,color,array}
\usepackage{marvosym}
\input{insbox}
\usepackage{graphicx}
\usepackage{wrapfig}
\usepackage{pifont}%
\usepackage{pgfplots}
\pgfplotsset{width=7cm,compat=1.8}
\usepackage{mdframed}
\usepackage{float,graphicx}
\usepackage{tikz}
\usetikzlibrary{arrows,shapes,fadings,positioning,shapes.geometric,backgrounds,
calc,decorations.pathmorphing}
\usepackage{lipsum}
\newcommand{\set}[1]{\{#1\}}    % set
\newcommand{\card}[1]{|{#1}|}   % cardinality of a set

\usepackage[noline,boxed,ruled,linesnumbered]{algorithm2e}
\usepackage[skins]{tcolorbox}

\makeatletter
\pgfdeclareshape{datastore}{
  \inheritsavedanchors[from=rectangle]
  \inheritanchorborder[from=rectangle]
  \inheritanchor[from=rectangle]{center}
  \inheritanchor[from=rectangle]{base}
  \inheritanchor[from=rectangle]{north}
  \inheritanchor[from=rectangle]{north east}
  \inheritanchor[from=rectangle]{east}
  \inheritanchor[from=rectangle]{south east}
  \inheritanchor[from=rectangle]{south}
  \inheritanchor[from=rectangle]{south west}
  \inheritanchor[from=rectangle]{west}
  \inheritanchor[from=rectangle]{north west}
  \backgroundpath{
    \southwest \pgf@xa=\pgf@x \pgf@ya=\pgf@y
    \northeast \pgf@xb=\pgf@x \pgf@yb=\pgf@y
    \pgfpathmoveto{\pgfpoint{\pgf@xa}{\pgf@ya}}
    \pgfpathlineto{\pgfpoint{\pgf@xb}{\pgf@ya}}
    \pgfpathmoveto{\pgfpoint{\pgf@xa}{\pgf@yb}}
    \pgfpathlineto{\pgfpoint{\pgf@xb}{\pgf@yb}}
 }
}
\makeatother
\begin{document}

\title{Dynamics and Allocation of Transaction Cost in Multiagent Industrial Symbiosis}

\titlerunning{Dynamics and Allocation of TC in Multiagent Industrial Symbiosis}

\author{Vahid Yazdanpanah\inst{1} \and
Devrim M. Yazan\inst{2} \and
W. Henk M. Zijm\inst{2}}

\authorrunning{V. Yazdanpanah et al.}

\institute{Agents, Interaction and Complexity Group \\
University of Southampton, Southampton, UK \\
\email{V.Yazdanpanah@soton.ac.uk}  \and 
Department of Industrial Engineering and Business Information Systems \\
University of Twente, Enschede, The Netherlands \\
\email{\{D.M.Yazan,W.H.M.Zijm\}@utwente.nl}}

\maketitle 

%======= Abstract =====================
\begin{abstract}
This paper discusses the dynamics of  Transaction Cost (TC)  in Industrial Symbiosis Institutions (ISI)  and provides a  fair and stable mechanism for TC allocation among the involved firms in a given ISI. In principle, industrial symbiosis, as an implementation of the circular economy paradigm in the context of industrial relation, is a practice aiming at reducing the material/energy footprint of the firm. The well-engineered form of this practice is proved to decrease the transaction costs at a collective level. This can  be achieved using information systems for: identifying potential synergies, evaluating mutually beneficial ones, implementing the contracts, and governing the behavior of the established relations. Then the question is ``\emph{how to distribute the costs for maintaining such an information system in a fair and stable manner?}'' We see such a cost as a  \emph{collective transaction cost} and employ an integrated method rooted in cooperative game theory and multiagent systems research to develop a fair and stable allocation mechanism for it. The novelty is twofold: in developing analytical multiagent  methods for capturing the dynamics of transaction costs in industrial symbiosis and in presenting a novel game-theoretic mechanism for its allocation in  industrial symbiosis institutions. While the former contributes to the theories of industrial symbiosis (methodological contribution), the latter supports decision makers aiming to specify fair and stable industrial symbiosis contracts (practical contribution). 
\end{abstract}

%==== BODY of the Paper =================

\section{Introduction}

Industrial symbiosis is a transitional business model to shift from linear economy paradigms towards implementing  the concept of circular economy---in the context of industrial relations/networks. In principle, the aim is to facilitate the circulation of reusable  resources among the network members~\cite{chertow2007uncovering,gonela2015stochastic}. Realizing such a form of collaboration requires methods for identifying potential matches~\cite{van2018literature}, evaluating them to generate mutually beneficial instances~\cite{yazdanpanah2019fisof}, implementing the cost-sharing schemes in bilateral contracts~\cite{IESM}, and decentralized governance of the established relations~\cite{yazdanpanah2016normative}. As elaborated in~\cite{DBLP:conf/atal/YazdanpanahYZ18} problematic situations occur when we move from bilateral relations to multilateral forms in multiagent industrial symbiosis. Dealing with such problems requires:  (1) developing practical methods able to capture collective-level concepts, such as  collectively realized transaction costs, (2) practice-oriented semantics to reason about the link between individual (firm-level) concerns and the collective (institution-level) attributes, and (3) implementable mechanisms to guarantee desirable collective attributes with no harm to firm-level concerns. This work aims to address this gap by focusing on the nature of collective transaction costs in industrial symbiosis and developing a contextualized allocation mechanism that guarantees game-theoretical fairness and stability properties.

\subsection{Conceptual Analysis: On the Nature and Dynamics of TC in IS}

As originally introduced by~\cite{williamson1979transaction}, Transaction Costs (TC) play a key role in the establishment and stability of different forms of contractual relations. This is also  the case  for industrial symbiosis relations, both in  bilateral and multilateral forms. For instance, when an industrial cluster manager aims to share  the collective costs for maintaining the shared environments  among the cluster members, she should take into account the firm-level concerns. One general approach is to see the contribution of each firm to the collective as a measure for making such cost/benefit allocation decisions~\cite{billera1982allocation}. In the case of IS, while we have such contribution-aware tools for allocating costs in  bilateral IS relations~\cite{IESM}, they  fail to guarantee some basic properties such as collective rationality (related to whether firms have any incentive to leave the collaboration) on a  multilateral network level~\cite{DBLP:conf/atal/YazdanpanahYZ18}. In practice, such a disadvantage  results in inefficient  deployment of an IS management platform in real-life industrial symbiosis networks.
 
To develop an efficient technique for dealing with this problem, it is crucial to understand the context and see what are transaction costs in IS.  In multiagent industrial symbiosis, the main elements that contribute to the transaction cost are costs involved in market/partner searching, negotiation costs, and the relation enforcement cost~\cite{fraccascia2018role,chertow2012organizing}. We argue that  in modern industrial symbiosis, and thanks to (online) IS information systems/platforms, the total IS transaction cost boils down to costs for establishment and maintenance of the information system which is responsible for handling the searching/matching process, for supporting/automating the negotiations,  and for synthesizing the required enforcement measures. Then the question is about finding methods to distribute this total cost among the involved firms, such that it would be fair with respect to each firm's contribution. That means that, after a basic payment for getting involved in a platform, firms that may gain more in a given network, are expected to  pay more for upcoming  transaction costs. In other words: ``\emph{with more power comes more responsibility}''. This perspective supports the so called  \emph{fairness} notions---proposed in cooperative game theory---that agents' individual benefit/cost share oughts to reflect their contribution to the collective  benefit/cost~\cite{deng1994complexity,young1985cost}. 
While the idea to take into account each agent's contribution provides a basis for allocating the costs in multiagent industrial symbiosis, we lack methods for defining the value of each and every coalition of firms, involved in the network\footnote{In the game-theoretic language, the characteristic function of the cooperative multiagent industrial symbiosis is not well-defined with respect to the context of industrial symbiosis and its constraints.}. Such an input is  crucial for applying standard fair allocation mechanisms (e.g., the notion of the Shapley value~\cite{shapley1953value}). In response, developing a method that characterizes  IS as a game is  the first objective of this work. The result of this first step will  be a game-form that in turn enables the application of game theoretic solution concepts for allocating the transaction cost. Technical details will follow in Section \ref{sec:modeling}.

To formulate a realistic game, based on which the transaction cost can be allocated, one question that is crucial to address is ``\textit{what is the nature of  transaction cost in IS practices?}''. In the IS literature~\cite{fraccascia2018role,yazdanpanah2019fisof}, costs for operationalizing IS are categorized as costs for transporting the resource, for treating them by means of recycling/preparation processes, and finally transaction costs as discussed above. In principle, transportation and treatment costs are based on the realization of physical facts and interactions\footnote{To reason about the essence of transaction costs, we use the terminology of~\cite{searle2005institution} and form our perspective based on  the categorization of \emph{physical} and \emph{institutional} facts and their corresponding acts. In brief, physical facts are about the valuation of a variable in the observable world (e.g., the \emph{quantity} of a resource or the \emph{distance} between two firms) while institutional facts are about the invisible world of concepts, definable in a given context (e.g., the fact that a resource \emph{is needed} or that a firm \emph{is powerful}). In turn, physical acts may change the value of physical facts while institutional acts affect the institutional facts. We abstract from the reasoning structures that relate the two disjoint sets of facts/acts. See~\cite{yazdanpanah2016normative} for further elaborations on such a conceptualization in the context of industrial symbiosis.}. For instance, the physical distance between the firms is one of the  key  measures that determines the transportation cost and hence can be a fair basis for allocating the collective transportation cost among the involved firms (e.g., see~\cite{sun2015transportation}). For treatment costs---as the total cost that results from recycling, drying, and categorizing---we also have physical processes that consume  resources (e.g., electricity and diesel cost of a recycling plant). However, the transaction cost is not  based on  physical facts, but mainly is the aggregation of costs of  institutional acts, e.g., negotiation costs and corresponding costs for establishing monitoring/enforcement mechanisms. These acts are on the one  hand required to enable or keep track of physical processes (hence have a  \emph{physical} connotation) and on the other hand  are related to the structure of interfirm connections (hence have an \emph{institutional} nature). Therefore, it  would be reasonable to consider a form of dual-nature \emph{physical-institutional IS value} as a basis for allocating the collective transaction cost among the involved firms. This way of capturing the institutional dimension of  the transaction cost  is in-line with the original studies on this notion, as presented in~\cite{williamson1993transaction,williamson1979transaction,silverman1999technological}. Although the line of reasoning is straightforward, capturing such a perspective in a practical tool for allocating the collective transaction costs in multiagent industrial symbiosis is an open problem\footnote{As highlighted by~\cite{DBLP:conf/atal/YazdanpanahYZ18}, some standard allocation mechanisms are inapplicable for IS implementation due to operational complexities---embedded in  real-life multiagent industrial relations.}.  Next, we elaborate on our approach on tailoring  formal multiagent  methodologies~\cite{wooldridge2013incentive,dawid2018agent} for modeling industrial symbiosis as an institution and for developing an operationally feasible transaction cost allocation mechanism.

\subsection{Institutional Approach: On How to Address the Problem}

We see multiagent industrial symbiosis as a practical manifestation of (well-designed) industrial institutions and aim to model it using game-theoretic methods, able to capture both the physical and the institutional contributions of involved firms\footnote{We use the term \emph{institution} as a general reference to a collective of entities behaving in a systematic manner, under an emerged or established coordination mechanism. Then a well-designed institution is one in which the mechanism is engineered such that some (collectively desirable) properties hold~\cite{rubino2005computational}. This would be distinguishable from the stronger notion of \emph{organization} where we have an explicit representation of \emph{roles}, organizational \emph{goals/preferences},  organizational \emph{structures}, and interaction \emph{protocols}~\cite{baldoni2010behavior,van2005formal}. }. Such a formal approach enables employing institutional economics methods for guaranteeing desirable properties at the collective level, e.g., see~\cite{DBLP:conf/atal/YazdanpanahYZ18} for how a regulatory agent can influence the feasibility of multiagent industrial symbiosis by means of incentive engineering techniques. 

In relation to the focus of this work, i.e., the collective transaction cost and its allocation among the involved firms, the  \emph{fairness} property is concerned with capturing the contributions of firms as a basis for cost allocation. While physical acts/facts (e.g., distance and quantifiable used energy) determine a base for computing fair allocation of (physical)  transportation/treatment costs, a fair  allocation of the collective transaction cost calls for a notion to capture the institutional contribution of agents as well.  To that end, we provide a formal account of the transaction cost economics in industrial symbiosis (based on  computational organization theory and industrial institutions). This results in the introduction of the notion of \emph{industrial symbiosis index} as a measure for capturing the physical and institutional contribution of firms. In turn, this notion will be a base for developing a fair and stable transaction cost allocation mechanism---rooted in the literature on fair division mechanisms~\cite{thomson1983fair,pratt1990fair}. Finally, we elaborate on potential questions to be solved using the provided methodological foundation.

\section{Modeling Multiagent Industrial Symbiosis Institutions}\label{sec:modeling}

To model IS institutions, we build upon a graphical representation of cooperative games---also known as \emph{graphical} games~\cite{kearns2008graphical} or \emph{graph-restricted} games~\cite{myerson1977graphs}. Such a representation is a natural choice as it reflects the established relations among the firms and allows the application of standard fair division methods for sharing the collective transaction cost, among the members of the institution\footnote{Through the course of this work, we may refer to firms as  \emph{agents}---following the convention in computational economics. This would be  to see any industrial symbiosis institution as an environment that supports the collaborative interaction of a set of autonomous, rational decision-makers in charge of the involved firms.}. As a first step, we use  graph-theoretic notions to determine a realistic characteristic function for the game-theoretic representation of industrial symbiosis. Then, adding an allocation mechanism results in our formal notion of industrial symbiosis institution.

\subsection{Preliminary Notions and Definitions}

We recall basic game theoretic notions and the definition of graphical games based on~\cite{mas1995microeconomic,myerson1977graphs}.

\textit{Cooperative Games:} A (transferable utility) cooperative game on a finite set of agents $\Gamma$ is a tuple $\coop{\Gamma,f}$ where the game's characteristic function $f: 2^{\Gamma} \mapsto \mathbb{R}$ is such that $f(\emptyset)=0$.

\textit{Graphical Games:}  A graphical (transferable utility) cooperative game on a finite set of agents/vertices $\Gamma$ is a triple $\coop{\Gamma,W,f}$ where $W$ is a $|\Gamma| \times |\Gamma|$ real-valued weight matrix (representing the weights of edges between vertices in $\Gamma$) and  the game's $W$-restricted characteristic function $f^W: 2^{\Gamma}  \mapsto \mathbb{R}$ is such that $f(\emptyset)=0$. We say $f$ is restricted to $W$ as it determines the value of any coalition $S \subseteq \Gamma\setminus\emptyset$ with respect to $W$. Such a general formalization allows further tailoring in the context of industrial symbiosis.

\textit{Allocation Mechanisms:} For a given cooperative game $\mathcal{G}=\coop{\Gamma,f}$, a (single-point) allocation mechanism $\mathcal{M}$ maps a real-valued tuple $\mathcal{M}(\mathcal{G}) \in \mathbb{R}^{|\Gamma|}$ to the pointed game. The $i$-th element of  the allocation tuple $\mathcal{M}(\mathcal{G})=\coop{a_1,a_2, \dots, a_{|\Gamma|}}$ is the \emph{share} of agent $i \in \Gamma$  according to $\mathcal{M}$ and with respect to  $\mathcal{G}$. The term \emph{share} can be interpreted---with respect to the context---as the amount to be  \emph{paid} or  \emph{gained} by $i$. We later discuss various properties that such a mechanism can hold or bring about.  

%====================================
\subsection{From Weighted Connectivity Graphs to Cooperative Graph Games}
%====================================

To determine how the transaction cost can be allocated among the firms based on their physical and institutional contributions, we take the graph that represents the established symbiotic relations and obtained cost reductions as an input\footnote{Note that the reasoning about such a cost allocation takes place in a retrospective manner and  after the establishment of industrial symbiosis relations.}. Given such a graph, we formulate a game-theoretical representation  that in turn results in inducing the physical as well as the institutional contribution of individual firms.  

\begin{definition}[IS Graph] An IS graph is a tuple $\coop{\Gamma,W}$, where  $\Gamma$ is the set of vertices, representing $|\Gamma|$ firms  and $W$ is the $|\Gamma| \times |\Gamma|$ matrix of positive  real valued  weights, representing the cost reduction values. There exists  a weighted undirected edge between distinct firms $i,j \in \Gamma$, representing their established symbiotic relation,  only if $W_{i,j} \neq 0$. Moreover, for any $i \in \Gamma$ we have that $\sum\limits_{j \in \Gamma} W_{i,j} > 0 $ (connected) and that $W_{i,i}=0$ (loop-free).
\end{definition}

To have a concise and contextualized representation, we don't require an explicit set of edges as it could be derived based on $W$. The same holds for requiring the graph to be loop-free and connected. Basically, loops and disconnected firms can be excluded as in such cases the transaction cost (hence its allocation) is meaningless. This results in  a realistic representation in which unfeasible relations/edges (which otherwise could be represented by negative or zero weights) are excluded. In the context of IS, $W_{i,j}$ reflects the realized net benefit---in terms of collectively obtained cost reductions---of the symbiotic relation between firms $i$ and $j$ on a given (quantity of) resource $r$. As discussed in~\cite{yazdanpanah2019fisof}, such a collective benefit can be computed by deducing the total operational cost of the relation (for treatment and transportation of $r$) from the total traditional costs (for discharging $r$ on the provider side of the relation and purchasing traditionally-used inputs---substituted by $r$ in the realized relation---on the receiver side). The $W$ graph would be the basis for formulating both the physical IS game (reflecting obtainable benefits) and the institutional game (modeling the institutional power of firms in the cluster).  

\begin{definition}[Physical IS Game] A graphical physical IS game is a triple $\coop{\Gamma,W,v}$, where $G=\coop{\Gamma,W}$ is an IS graph and for any group of firms $S \subseteq \Gamma$ with $|S|>1$, the characteristic function $v(S)$ is equal to $\frac{1}{2}  \sum_{i,j \in S} W_{i,j}$. By convention, for any  $S$ with $|S|\leq 1$, $v(S)=0$. Then in the normalized characteristic function, denoted by $\bar{v}$, we have that $\bar{v}(S)=v(S)/v(\Gamma)$.
\end{definition}

\begin{example} \label{ex:case} To demonstrate the applicability of our approach, we use a case study (adopted from a realistic industrial cluster\footnote{The adopted case is one of the successful implementations of industrial symbiosis networks, investigated in the a European project. Due to confidentiality concerns, we omitted the company names and modified some values.}).  See Figure \ref{fig:graph} for an illustration of the IS graph. In this graph, the value on each edge reflects the  benefit (in terms of cost reductions) that resulted from the symbiotic relation, realized between the nodes that it connects. While such values represent the physical dimension of an IS practice, the structure of the graph is what we later use to formulate the institutional importance of each node/firm.

%================ begin Graph ==============
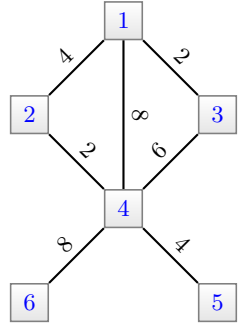
\begin{figure}[!htb] \label{fig:graph}
\centering 
\begin {tikzpicture}[-latex ,auto ,node distance =1.25cm and 1.25cm ,on grid ,
semithick,
state/.style ={ rectangle ,top color =white , bottom color = gray!20 ,
draw,gray , text=blue , minimum width =.5 cm, minimum height =.5 cm}]

\node[state] (1) {$1$};
\node[state] (2) [below left=of 1] {$2$};
\node[state] (3) [below right =of 1] {$3$};
\node[state] (4) [below right =of 2] {$4$};
\node[state] (5) [below right =of 4] {$5$};
\node[state] (6) [below left =of 4] {$6$};

\path[-, draw,thick]
    (1) edge [pos=0.5, sloped, above] node {$4$} (2)
    (1) edge [pos=0.5, sloped, above] node {$2$} (3)
    (1) edge [pos=0.5, sloped, above] node {$8$} (4)
    (2) edge [pos=0.5, sloped, above] node {$2$} (4)
    (3) edge [pos=0.5, sloped, above] node {$6$} (4)
    (4) edge [pos=0.5, sloped, above] node {$4$} (5)
    (4) edge [pos=0.5, sloped, above] node {$8$} (6)
    ;
\end{tikzpicture}
\caption{Connectivity graph of the involved firms in the IS cluster. Each node represents a firm and the value on each edge represents the amount of cost reduction, obtained as a result of the collaboration between the firms that the edge connects. Values are presented as \emph{utils}. A util can be interpreted as any form of transferable utility, e.g., a util may be equal to $100 \$$. } 
\end{figure}
%================ end Graph ==============

The value of any singleton or empty coalition is $0$ while for any $S$ with $|S| > 1$, we calculate the value by simply adding up the weights  of the edges that connect any two member of $S$. For instance, $v(\set{1,2,3,4})=22$ and $v(\Gamma=\set{1,\dots, 6})=34$. For the same coalitions, the normalized values are respectively $\frac{22}{34}$ and $\frac{34}{34}$. These normalized values will be later employed for aggregation of the physical game with a game that represents the institutional power of firms in such clusters. 
\end{example}

Below, we present game theoretic properties of the physical IS games.   

\begin{proposition}[Properties] \label{prop:1} Let $\mathcal{G}=\coop{\Gamma,W,v}$ be a physical IS game. Then: 
\begin{enumerate}[(1)]
    \item for any coalitions $S \subset T$, we have that $v(S) \leq v(T)$ (monotonicity).
    \item for any disjoint coalitions $S$ and $T$, we have that  $v(S \cup T) \geq v(S) + v(T)$ (super-additivity).
    \item for any coalitions $S$ and $T$, we have that $v(S \cup T) + v(S \cap T) \geq v(S) + v(T)$ (convexity/super-modularity).
\end{enumerate}

\end{proposition}
\proof
\textit{(1)} imagine a firm $i$ in $T\setminus S$. If $i$ is connected to a member of $S$ or $T \setminus S$, it contributes to the value of $T$. Otherwise, it has the added value of $0$. In both cases, part (1) is true. 
\textit{(2)} if there exist a direct edge connecting a member of $S$ to a member of the disjoint coalition  $T$, then $v(S \cup T)$ increases; otherwise, it is equal to $v(S) + v(T)$, thanks to non-negative weights. 
\textit{(3)} In case the two sets are disjoint, it follows from part (2). Otherwise, the two coalitions have nonempty intersection with either positive or zero value.\qed\endproof

These properties result in the following practical result that the collective value of the grand coalition $\Gamma$ can be shared among the firms such that no coalition has an incentive to defect the collaboration.

\begin{theorem}[Nonempty Core] \label{thereom:core_of_pysical} Let $\mathcal{G}=\coop{\Gamma,W,v}$ be a physical IS game. Then there exists an allocation mechanism $\mathcal{M}$ such that in $\mathcal{M}(\mathcal{G})=\coop{a_1,a_2, \dots, a_{|\Gamma|}}$, we have  (1) $\sum\limits_{i \in \Gamma} a_i = v(\Gamma)$ (Efficiency) and (2) for any $S \subseteq \Gamma$, $\sum\limits_{i \in S} a_i \geq v(S)$ (Coalitional Rationality). 
\end{theorem}
\proof The two clauses in this theorem axiomatize the notion of core-nonemptiness~\cite{shapley1969market}. Then, the convexity property (in Proposition \ref{prop:1}) in combination with the well-established Bondareva-Shapley theorem~\cite{bondareva1963some,shapley1967balanced}  ensures the nonemptiness of the core, and accordingly the existence of a mechanism to generate the allocation.\qed\endproof

\subsection{From Cooperative Games to Multiagent Industrial Institutions}

In addition to the physical game---based on which we can induce the \emph{physical} contributions that firms can bring about through a  bilateral exchange of resources---the next step is to present a basis for capturing the \emph{institutional} contribution of firms. For such a purpose, we cannot rely on cost reduction values (obtained thanks to operationalizing the relations) basically because transaction costs are non-operational, but have an institutional nature. To that end, we employ  the notion of \emph{closeness centrality} adopted from the literature on communication networks~\cite{bavelas1950communication} to capture the institutional power of firms, and as the basis for defining the characteristic function of the IS \emph{institutional} game. In a graph on the set of vertices $\Gamma$, the closeness centrality of a vertex $i \in \Gamma$, denoted by $\mathfrak{C}(i)$, is equal to $\frac{|\Gamma|-1}{\sum\limits_{j \in \Gamma\setminus\set{i}}   d (i,j)}$ where the distance function $d: \Gamma \times \Gamma \mapsto \mathbb{N}^+$ returns the shortest distance between $i$ and $j$. More explicitly,  $d(i,j)=d(j,i)$ returns  the minimum number of edges  passed to reach $j$ from $i$. Recalling the presented IS graph in Figure \ref{fig:graph}, we have that $\mathfrak{C}(1)=\frac{5}{7}$, $\mathfrak{C}(2)=\mathfrak{C}(3)=\frac{5}{8}$, $\mathfrak{C}(4)=\frac{5}{5}$, and $\mathfrak{C}(5)=\mathfrak{C}(6)=\frac{5}{9}$. 

Then based on this notions, we formulate the institutional IS game as follows. 

\begin{definition}[Institutional IS Game] A graphical institutional IS game is a triple $\coop{\Gamma,W,\iota}$, where $G=\coop{\Gamma,W}$ is an IS graph and for any group of firms $S \subseteq \Gamma$ with $|S|>0$, the characteristic function $\iota(S)$ is equal to $\sum_{i \in S} \mathfrak{C}(i)$. By convention, $\iota(\emptyset)=0$. Then in the normalized characteristic function, denoted by $\bar{\iota}$, we have that $\bar{\iota}(S)=\iota(S)/\iota(\Gamma)$.
\end{definition}

Due to the additive formulation of $\iota$, we have the following properties for institutional IS games.

\begin{proposition}[Properties] \label{prop:2} Let $\mathcal{G}=\coop{\Gamma,W,\iota}$ be an institutional IS game. Then $\mathcal{G}$ is (1) monotonic and (2) convex/super-modular (defined analogously to Proposition \ref{prop:1}). 
\end{proposition}
\proof Given that $\mathfrak{C}(i)$ is non-negative for all $i \in \Gamma$ and the formulation of $\iota(S)$ as the summation of $\mathfrak{C}(i)$ for all $i \in S$, monotonicity is trivial. For   convexity, it suffices to decompose $\iota(S \cup T)$. We have that $\iota(S \cup T)$ is equal to $\sum_{i \in S \cup T} \mathfrak{C}(i) = \sum_{j \in S } \mathfrak{C}(j) + \sum_{k \in T } \mathfrak{C}(k) - \sum_{l \in S \cap T } \mathfrak{C}(l)$, hence  $\iota(S \cup T) + \iota(S \cap T) = \iota(S) + \iota (T)$,  which satisfies the convexity condition.\qed\endproof

Immediate to this, we have the existence of an efficient and coalitionally rational allocation mechanism for any institutional IS game (parallel to Theorem \ref{thereom:core_of_pysical} for physical IS games). 

Having both the physical and the institutional aspects of IS formalized in the game-theoretic language, we present the aggregated IS game as the summation of the normalized  form of the two games. 

\begin{definition}[IS Game] Let $G=\coop{\Gamma,W}$ be an IS graph,  $\mathcal{G}_P=\coop{\Gamma,W,v}$ a physical IS game on $G$, and $\mathcal{G}_I=\coop{\Gamma,W,\iota}$ an institutional IS game on $G$. Then the graphical IS game is a triple $\coop{\Gamma,W,\sigma}$, such that for any group of firms $S \subseteq \Gamma$, the characteristic function $\sigma(S)$ is equal to $\bar{v}(S)+\bar{\iota}(S)$. 
\end{definition}

Then, as a corollary to Propositions \ref{prop:1} and \ref{prop:2} and Theorem \ref{thereom:core_of_pysical},  we immediately deduce  that any IS game preserves the properties presented in Proposition \ref{prop:2}, hence has a non-empty core (analogous to Theorem \ref{thereom:core_of_pysical}).

\begin{corollary} \label{corollary} For any IS game $\coop{\Gamma,W,\sigma}$, the set of efficient and coalitionally rational allocation mechanisms is non-empty. 
\end{corollary}

In other words: the normalized versions of both games satisfy the presented properties and linear aggregation preserves them. 
In some application domains, one may opt for various forms of linear aggregations, i.e., to employ $\sigma= \alpha \bar{v} + \beta \bar{\iota}$ (for integer-valued positive $\alpha$ and $\beta$). We later highlight that due to the linearity of the allocation mechanism that we employ, our results remain  valid in such a generalization of the problem.

Following the presented perspective in~\cite{yazdanpanah2019fisof}, an industrial symbiosis institution consists of a group of firms, a structure that specifies the outcome of collaboration among potential coalitions (in the group), and mechanism(s) responsible for coordinating the institution. Such mechanisms are basically  in charge to guarantee some desirable properties in the institution. In our case---in industrial symbiosis institutions---the aim could be to ensure the \emph{stability} of the institution (i.e., that no firm or group of firms has an incentive to defect the collaboratively profitable institution), the \emph{fair allocation} of the collectively obtained benefits (such that the contribution of firms is reasonably  reflected in their individual shares), or ideally to bring about both \emph{fairness and stability}. In a general form, an industrial symbiosis institution is defined as:

\begin{definition}[Industrial Symbiosis Institution] Let $\Gamma$ be a set of firms, $\mathcal{G}$ an IS game among firms in $\Gamma$, and $\mathfrak{M}$ a set of value allocation mechanisms. Then an  industrial symbiosis institution is defined as triple $\mathcal{I}=\coop{\Gamma,\mathcal{G},\mathfrak{M}}$. 
\end{definition}

In brief, this is to see an IS institution  as an IS game under mechanisms in charge of distributing the collective values\footnote{Note that we do not fix the allocation mechanism but take a set $\mathfrak{M}$.}. This would be to distribute collectively obtainable benefits as well as collective operational costs for establishment and maintenance of the institution. The latter category corresponds to the focus of this work for allocating transaction costs in industrial symbiosis. In the next section, we present an allocation mechanism---corresponding to the notion of Shapley value and the Myerson value in graph-restricted games~\cite{shapley1953value,myerson1977graphs}---that satisfies both \emph{fairness} and \emph{stability} properties. We also elaborate on its computational complexity and tractability results.

\section{A Shapley-Based Allocation Mechanism for IS}

The main idea behind the \emph{fair} allocation of values is to take into account the contribution of each agent to the collaborative group~\cite{mas1995microeconomic}. In industrial symbiosis management platforms, the \emph{collective}  transaction cost reduces to costs for establishment and maintenance of the framework---as a dynamic e-market environment\footnote{Note that there may exist  other forms of \emph{individual} transaction costs in such relations. For instance, some investments to calibrate the production process in order to enable accepting a waste-based resource would be categorized as a transaction cost. However, as the equipment remains property of the firm---and can be potentially used  in its further relations---it is not reasonable to consider such a cost as a collective (to be shared) transaction cost.}. This calls for dynamic cost allocation methods, able to grasp the physical as well as institutional nature of each  agent's contribution. Roughly speaking, following an initial payment for a basic membership (to get involved in this e-market framework) it is expected that a ``\emph{fair}'' allocation of costs for further improvements, takes into account the contribution of each participant---in terms of their role/function in the formed industrial network. 
Having the game-theoretic formulation of an IS game (as an aggregation of the physical and institutional IS games), any proportion of the transaction cost that oughts to be shared among the firms---e.g., the total cost for updating the IT platform---can be distributed based on each firm's contribution to the IS game. A standard notion in computational economics, capturing the contribution of each agent in a cooperative  setting, is the \textit{Shapley value}~\cite{shapley1953value}. Shapley's allocation method uniquely satisfies the so called \emph{fairness} properties, which has high relevance for our domain of application in industrial organizations.  In IS games, as a combination of normalized physical and institutional games among the firms, the Shapley value of a firm  determines the extent of its power and influence in the institution. This value would be defined as what we call the firm's \emph{IS index}. 

\begin{definition}[IS Index] Let $\mathcal{G}=\coop{\Gamma,W,\sigma}$ be an IS game. Then for any arbitrary $i \in \Gamma$, the IS index, denoted by $\Phi_i(\sigma)$, is equal to  $\sum\limits_{S \subseteq \Gamma \setminus \set{i}} \frac{\card{S}! (\card{\Gamma}-\card{S}-1)!}{\card{\Gamma}!} (\sigma(S \cup \set{i})-\sigma(S))$. \end{definition}

Due to the characteristics of $\sigma$ (as a reflection of both the physical as well as institutional aspects of IS) the introduced notion of \emph{IS index} is a measure that reflects the power of a firm based on  its connectivity to other firms in the network and also its operational contributions by bringing about cost reductions\footnote{We later show that  due to the graphical representation of the problem, the IS index can be formulated in a non-factorial manner. This results in lower computational complexity for calculating the IS index of each firm.}. 
This index forms  a basis for allocating transaction costs such that a higher contribution determines a higher share (i.e., \emph{``with more power comes more responsibility''}). This approach relies on the standard rationale in cooperative cost-sharing  games that agents with higher potentials ought to pay the larger share of the costs in the  collaborative practice~\cite{marden2013distributed}. Accordingly, given an external cost function $\tau(\Gamma)$ (equal to $\tau(S)$ for all $S \subseteq \Gamma$ with $|S|\geq 2$),  determining the (to be shared) transaction cost\footnote{ Such a cost may consist of initial platform development costs, ongoing IT infrastructure maintenance, or extra personnel recruitment costs for updating the platform. $\tau$ is defined in a functional way merely to allow further extensions on dynamic formulations of the transaction cost function, e.g., as a temporal function of some required resources for maintaining an IS information system.  In other words, $\tau(S)$ is undefined for $|S|\leq 1$ and is equal to a given value $\tau(\Gamma)$ otherwise. Note that our main question is on how to distribute this collectively defined value.} among the members of  $\Gamma$, we present the following transaction cost allocation mechanism for IS institutions.

\begin{definition}[TC Allocation for IS] \label{def:allocation_mech} Let $\mathcal{G}=\coop{\Gamma,W,\sigma}$ be an IS game, $\Phi_i(\sigma)$ the corresponding IS index for any arbitrary $i \in \Gamma$, and $\tau(\Gamma)$  a given transaction cost value for $\Gamma$. We define the cost share of agent $i \in \Gamma$ as $T_i(\sigma, \tau):=  \frac{\Phi_i(\sigma) \cdot \tau(\Gamma)}{\sigma(\Gamma)}$.  Allocation $T(\sigma,\tau)=\coop{a_1,    \dots, a_{|\Gamma|} }$ with  $a_i=T_i(\sigma, \tau)$ denotes the TC allocation for IS game $\mathcal{G}$ with respect to $\tau$. \end{definition}

This allocation, tailored and contextualized for the specific  class of structured, graph-restricted  industrial symbiosis games, (1) captures both the physical as well as the institutional aspects of this practice, (2) satisfies desirable fairness and stability properties (to be discussed in Section \ref{sec:fairness}), and (3) is computationally tractable thanks to the graphical  representation of the games (to be illustrated in Section \ref{sec:complexity}).  

%======================
\subsection{On Fairness and Stability} \label{sec:fairness}
%======================

Having an industrial institution, \emph{stability} and \emph{fairness} are  two properties insurable by means of well-designed mechanisms.  In the case of the transaction cost allocation mechanism, stability is about (1) sharing the exact amount of the cost and (2) sharing such that no firm can benefit by defecting from the institution. On the other hand, fairness is a more complex property, concerned with (1) the symmetric contribution of firms to the institution, (2) the share of firms whose involvement are non-contributory, (3) the possibility to aggregate various institutions, and finally (4) the sharing of the  exact total cost. Below, we provide a formal account of these properties in axiomatic forms---based on~\cite{mas1995microeconomic}---and  investigate whether they are valid in case of our suggested cost allocation mechanism. 

\begin{proposition}[Fairness Axioms] Let $\mathcal{I}=\coop{\Gamma,\mathcal{G},\mathfrak{M}}$ be an industrial symbiosis institution where $\Gamma$ is the set of firms and $\mathcal{G}=\coop{\Gamma,W,\sigma}$ is the IS game. For any transaction cost  $\tau(\Gamma)>0$ we have that  $\mathfrak{M}=\set{T(\sigma,\tau)}$ (Definition \ref{def:allocation_mech}) guarantees the following fairness axioms:
\begin{enumerate}[(1)]
    \item The collective transaction cost is efficiently allocated among the firms, formally,  $\sum\limits_{i \in \Gamma} T_i(\sigma,\tau) = \tau(\Gamma)$ (Efficiency).
    \item The  identities of the  firms do  not  affect  their share of the total transaction cost, formally, for $i,j \in \Gamma$, $T_i(\sigma,\tau) = T_j(\sigma,\tau)$ if for all $S \subseteq \Gamma \setminus \set{i,j}$ we have that $\sigma (S \cup \set{i})= \sigma (S \cup \set{j})$ (Symmetry).
    \item Any firm of which its contribution to any coalition is equal to its individual value, pays a transaction  cost share proportional to its individual value, formally, for $i \in \Gamma$, $T_i(\sigma,\tau) = \frac{\sigma(\set{i}) \cdot \tau(\Gamma)}{\sigma(\Gamma)}$ if for all $S \subseteq \Gamma \setminus \set{i}$ we have that $\sigma (S \cup \set{i})= \sigma (S) + \sigma(\set{i})$(Dummy Player).
    \item For two IS games, an agent's transaction cost share in the  aggregated game is equal to the summation of its share in each, formally, given an industrial game $\mathcal{G}' = \coop{\Gamma,W', \sigma'}$ and a corresponding transaction cost $\tau'(\Gamma)>0$, we have that $T_i(\sigma+\sigma',\tau+\tau')= T_i(\sigma,\tau)+T_i(\sigma',\tau')$ (Additivity).
\end{enumerate}
\end{proposition}
\proof Our notion of IS index measures  the Shapley value of each firm $i$. Following the linearity of this Shapley-based value, we have that the allocation mechanism preserves all the  four properties that the Shapley value uniquely possesses~\cite{hart1989shapley}.\qed\endproof

In general, fairness and stability are orthogonal. In other words, an allocation might be fair but not stable or the other way around.  Below, we present an axiomatic account of stability and show their validity  for the presented transaction cost allocation method. 

\begin{proposition}[Stability Axioms] Let $\mathcal{I}=\coop{\Gamma,\mathcal{G},\mathfrak{M}}$ be an industrial symbiosis institution where $\Gamma$ is the set of firms and $\mathcal{G}=\coop{\Gamma,W,\sigma}$ is the IS game. For any transaction cost  $\tau(\Gamma)>0$ we have that $\mathfrak{M}=\set{T(\sigma,\tau)}$ (Definition \ref{def:allocation_mech}) guarantees the following stability axioms:
\begin{enumerate}[(1)]
    \item The collective transaction cost is efficiently allocated among the firms, formally,  $\sum\limits_{i \in \Gamma} T_i(\sigma,\tau) = \tau(\Gamma)$ (Efficiency).
    \item No subgroup faces an economic incentive to deviate from the grand coalition and benefit by paying a lower share of the transaction cost, formally, for any coalition $S \subseteq \Gamma$ with $|S|\geq 2$, we have that $\sum\limits_{i \in S} T_i(\sigma,\tau) \leq \tau(S)$ (Coalitional Rationality).
\end{enumerate}
\end{proposition}
\proof The first part is valid (using  the previous proposition). For the second part,  the current formulation of the collective transaction cost requires that $\tau(S)=\tau(\Gamma)$ which if combined with the first clause, immediately satisfies the claim\footnote{In the generalized form, where the transaction cost function is defined for all  potential coalitions, the convexity of $\tau$ would be required for coalitional rationality. Note that in case such a function is available,  the mere problem on ``\emph{how to distribute the collective cost among the firms}'' will evaporate, as its solution requires a single call to this function.}.\qed\endproof

Thanks to the adoption of a Shapley-based index---and its linearity property---the fairness and stability properties will be preserved in the presented aggregated form of IS institutions and any general linear aggregation forms in which  the importance of the physical and institutional contributions are weighted.

\begin{theorem}[Generalizability]  Let $\mathcal{I}=\coop{\Gamma,\mathcal{G},\mathfrak{M}}$ be an industrial symbiosis institution where $\sigma= \alpha \bar{v} + \beta \bar{\iota}$ is the characteristic function of $\mathcal{G}$ in terms of $v$ and $\iota$, the corresponding characteristic functions in the physical and institutional IS games, respectively.  We have that  $\mathfrak{M}=\set{T(\sigma,\tau)}$  (Definition \ref{def:allocation_mech}) guarantees  fairness and stability in $\mathcal{I}$.
\end{theorem}

%======================
\subsection{Reductions and a Tractable Algorithm} \label{sec:complexity}
%======================

Although the presented Shapley-based IS index has the above-mentioned desirable properties, its  standard formulation leads to computationally expensive algorithms. Below, we present reductions that result in an alternative formulation for computing the IS index. 

\begin{lemma} \label{lemma:red_phys} In a graphical physical IS game $\mathcal{G}_P=\coop{\Gamma,W,v}$, for any $i \in \Gamma$ we have that $\sum\limits_{S \subseteq \Gamma \setminus \set{i}} \frac{\card{S}! (\card{\Gamma}-\card{S}-1)!}{\card{\Gamma}!} (v(S \cup \set{i})-v(S))=\sum\limits_{j \in \Gamma\setminus\set{i}} \frac{W_{i,j}}{2}$. \end{lemma}
\proof Based on the formulation of $v$, we have that the value of any singleton coalition $S$ is zero and for any coalition $T$ with more than two members, the value is computed based on the summation of values that bilateral relations (established within $T$) bring about. In other words, the average marginal contribution of any firm to any $T$ with more than two members is zero. The only set of coalitions to which a firm may have a contribution are two-member coalitions for which we have the results of~\cite{IESM} that the middle point of the core corresponds to the average marginal contribution. In our graph-restricted games,  this value is equal to the Myerson value~\cite{myerson1977graphs} and is equal to half of the summation of the values on the edges that are directly connected to $i$, i.e., $\sum\limits_{j \in \Gamma\setminus\set{i}} \frac{W_{i,j}}{2}$.\qed\endproof

Next, we present a reduction for computing the contributions in the institutional game.

\begin{lemma} \label{lemma:red_inst} In a graphical institutional IS game $\mathcal{G}_I=\coop{\Gamma,W,\iota}$, for any $i \in \Gamma$ we have that $\sum\limits_{S \subseteq \Gamma \setminus \set{i}} \frac{\card{S}! (\card{\Gamma}-\card{S}-1)!}{\card{\Gamma}!} (\iota(S \cup \set{i})-\iota(S))=\mathfrak{C}(i)$.
\end{lemma}
\proof In the institutional game, each firm $i$'s contribution to any coalition is equal to its degree of closeness centrality. Then the dummy player property implies the claim.\qed\endproof

Based on these reductions we have that the transaction cost allocation is computationally tractable. 

\begin{theorem} Let $\mathcal{I}=\coop{\Gamma,\mathcal{G},\mathfrak{M}}$ be an industrial symbiosis institution where $\Gamma$ is the set of firms and $\mathcal{G}=\coop{\Gamma,W,\sigma}$ is the IS game. For any transaction cost  $\tau(\Gamma)>0$, employing $\mathfrak{M}=\set{T(\sigma,\tau)}$ to compute the allocation $T(\sigma,\tau)=\coop{a_1,\dots,a_{|\Gamma|}}$ is polynomial in time and space. \end{theorem}
\proof We present a constructive proof by providing an algorithm (see Algorithm \ref{alg:allocation}) that generates the allocation, of which we verify its correctness and subsequently prove the complexity claims.

%====================== ALG BEGIN ======================
\begin{algorithm}[!htb] \label{alg:allocation}

 \DontPrintSemicolon
 \SetKwFunction{SDud}{SDud}\SetKwFunction{Tol}{tol}\SetKwFunction{Break}{break}% tt
 \SetKwInOut{Input}{Input}
 \SetKwInOut{Output}{Output}
 \Input{IS Graph $G=\coop{\Gamma,W}$ with $\Gamma$ as the indexed set of firms and $W$ as the $|\Gamma| \times |\Gamma|$ weight matrix, $\:$ Transaction Cost $\tau(\Gamma)$.}
 \emph{\% Initialization}\;
 $n \leftarrow \card{\Gamma}$\;
 $Sum(G) \leftarrow \frac{1}{2} \sum\limits_{i,j \in \Gamma} W_{i,j}$  \;
 $Cent(G) \leftarrow \sum\limits_{i \in \Gamma} \mathfrak{C}(i)$\;
 $T \leftarrow [T_1,\dots,T_{n}]$ \emph{\% n-Member Allocation Array} \;
 \emph{\% Allocation}\;
 \For{$i \in \Gamma$}{ 
  $Sum(i) \leftarrow 0$\;
  \For{$j \in \Gamma\setminus\set{i}$}{ 
  $Sum (i) \leftarrow Sum(i) + \frac{W_{i,j}}{2}$\; 
  }
  \emph{\% Compute  IS Index $\Phi_i(\sigma)$}\;
  $\Phi_i(\bar{v}) \leftarrow  \frac{ Sum(i)}{Sum(G)}   $\;
  $\Phi_i(\bar{\iota}) \leftarrow  \frac{ \mathfrak{C}(i)}{Cent(G)}   $\;
  $\Phi_i(\sigma) \leftarrow \Phi_i(\bar{v}) + \Phi_i(\bar{\iota})     $ \;
  \emph{\% Compute Individual Transaction Cost $T_i(\sigma,\tau)$}\;
  $T_i(\sigma, \tau) \leftarrow  \frac{\Phi_i(\sigma) \cdot \tau(\Gamma)}{2}$\;
  $T[i] \leftarrow T_i(\sigma, \tau)$\;
}
return $T$\;

\caption{TC Cost Allocation in IS}
\end{algorithm}
%======================= ALG End ====================== 

 \textit{Correctness:} In Algorithm \ref{alg:allocation}, we have that for each firm $i \in \Gamma$, the IS index $\Phi_i(\sigma)$ is equal to the Shapley value of $i$ in the aggregated game (of the normalized physical and institutional games). Thanks to the additivity property, this would be equal to the aggregation of Shapley values in each game. Then, we rely on Lemma \ref{lemma:red_phys} and \ref{lemma:red_inst} for calculating the two values. Finally, for computing individual transaction costs, we have that $\sigma(\Gamma)=2$ as it is equal to $\frac{v(\Gamma)}{v(\Gamma)}+\frac{\iota(\Gamma)}{\iota(\Gamma)}$. 
\textit{Space Complexity:} The required matrix of weights (representing the set of obtained cost reductions) is in $O(n^2)$ where $n$ is the size of $\Gamma$. 
\textit{Time Complexity:} For computing the IS indices, we have $O(n)$ on the big loop. Then in the physical game component, $Sum(i)$ is in $O(n)$ (a pass on the $i$-th row in $W$) and $Sum(G)$ is in $O(n^2)$ (a pass through the whole $W$).  For the institutional part, we have that computing $\mathfrak{C}(i)$ is reducible to finding the  shortest paths~\cite{eppstein2001fast} which is well-known to be in $O(n^3)$~\cite{floyd1962algorithm}.\qed\endproof

\section{Method Application in a Case Study}

To show the applicability of the developed method  for allocating collective transaction costs among a cluster of firms $\Gamma$, we use the presented case in Example \ref{ex:case} and assume a total value $\tau(\Gamma)$ as the collective transaction cost, realized for an updating round in $\Gamma$'s industrial symbiosis information system. Assuming $\tau(\Gamma)=100$ simply results in percentage calculation for individual shares. 

Following the steps in Algorithm \ref{alg:allocation}, we have that $Sum(G)=34$ and $Cent(G)=\frac{1027}{252}$. Then for each firm $i \in \set{1,\dots,6}$, to compute $\Phi_i(\bar{v})$ (as the physical component of $\Phi_i(\sigma)$), we calculate the summation of the weights on all the edges connected to $i$ and divide it  by $Sum(G)$. Thus we have:  $\Phi_1(\bar{v})=\frac{7}{34}$, $\Phi_2(\bar{v})=\frac{3}{34}$, $\Phi_3(\bar{v})=\frac{4}{34}$, $\Phi_4(\bar{v})=\frac{14}{34}$, $\Phi_5(\bar{v})=\frac{2}{34}$, $\Phi_6(\bar{v})=\frac{4}{34}$. For each firm $i$, adding $ \frac{\mathfrak{C(i)}}{Cent(G)}$ to $\Phi_i(\bar{v})$ results in its IS index $\Phi_i(\sigma)= \frac{13309}{34918}, \frac{4218}{17459}, \frac{9463}{34918}, \frac{11473}{17459}, \frac{3407}{17459}, \frac{4434}{17459}$ (respectively for firms $1$ to $6$). Finally, the transaction cost allocation $T$ could be generated based on $\Phi_i(\sigma)$. We have that: $T_1(\sigma,\tau)= 19.06$, $T_2(\sigma,\tau)= 12.08$, $T_3(\sigma,\tau)= 13.55$, $T_4(\sigma,\tau)= 32.86$, $T_5(\sigma,\tau)= 9.76$, and $T_6(\sigma,\tau)= 12.70$. 

Note that as we employ generic graph-/game-theoretical solution concepts as a basis for the developed algorithm, our results are neither sensitive to  the  distribution of the cost reduction values nor to the structure of the connectivity graph.

\section{Conclusions and Open Research Directions}

Amid the institutional nature of transaction costs, to our knowledge, this work is the first proposal that translates Searle's well-established philosophy on institutional theory for the context of IS, takes it into practice for fair transaction cost allocation, and introduces a tractable algorithm for such a purpose. As a managerial decision support tool, the presented algorithm can be integrated into smart IS contracting and management frameworks to enable the automation of cost allocation procedures. For instance, as a suggested business model for IS clusters, firms would be expected to pay an initial membership fee and then be charged for further collective transaction costs based on the presented method---reflecting their operational as well as institutional contributions.

As presented, this work has immediate applicability to support IS management by means of providing a fair and stable TC allocation mechanism. In addition, it opens new research directions to study: different  \emph{IS classes}, various forms of \emph{TC functions}, applicable \emph{notions of fairness and toolbox of mechanisms}, and \emph{governance mechanisms} able to observe the collective behavior of IS.

\textit{IS Classes and Graphical Constraints:} While we focused on generic IS, an interesting line of research is to investigate sub-classes of IS with respect to their graphical structures. For instance, in most bio-based IS practices, bio-refineries are in the center of the cluster due to their crucial role as a resource treatment/recycling facility. This results in tree-like structures or in clusters of star graphs such that no two firms can implement an IS relation in the absence of a third-party refinery. This calls for tailored methods able to capture such contextual properties. To this end, a combination of tree-like graphical games~\cite{nisan2007algorithmic}  and  dependence graphs~\cite{DBLP:journals/cmot/ConteS02} would be a suggested formal foundation.

\textit{Transaction Cost Functions:} An immediate extension to this work is  to study transaction cost functions, formulated in terms of the operational dimensions of IS (see~\cite{yazdanpanah2019fisof} for such dimensions). This results in cases where---in addition to the grand coalition---the collective transaction cost for some (but not all the) other coalitions is well-defined,  through a functional formulation of TC; e.g., using the presented multiagent methods in~\cite{ping1996multi}. In such a case, we see that satisfying the coalitional rationality, by means of an allocation mechanism, calls for integrating the properties of such a function and requiring circumscriptions on the characteristic functions of the physical and institutional IS games.

\textit{Notions of Fairness and Toolbox of Mechanisms:} We plan to relate the presented Shapley-based fair allocation  to other axiomatic notions of fairness~\cite{axiomatic_fairness,richards2016personalized}. This results in a toolbox of mechanisms, each satisfying a set of desirable properties, hence admissible in a related domain of application. In practice, implementing such a mechanism toolbox in IS information systems enables the firm managers to opt  for a relevant set of decision support services, tailored with respect to their managerial strategies and preferences.    To compare the effectiveness of mechanisms in different domains, an appropriate approach would be to employ agent-based simulation techniques~\cite{abm1IEEE,abm2BSE} in combination with  participatory policy analysis tools \cite{mehryar2019individual,mehryar2020making}.

\textit{Norm-Aware IS Governance:} Finally, a line of focus for future work is to develop governance frameworks for IS. This is to enable monitoring of the organizational behavior and enforcing normatively desirable behaviors. For such  a  purpose, we  rely on the rich literature on norm-aware coordination~\cite{singh2013norms,dastani2017commitments,sikora1998multi} and address open problems related to \emph{organizational characterization}~\cite{DBLP:conf/atal/CoutinhoBSHB09} of multiagent industrial symbiosis. To that end, cost allocation and incentive engineering techniques can be employed as a financial instrument for nudging the collective behavior and to ensure the compliance of firms to contractual commitments~\cite{anderson2005management,chalioti2017strategic}. 

\subsection*{Acknowledgements} \emph{SHAREBOX}~\cite{sharebox}, the project leading to this work, has received funding from the European Union's Horizon 2020 research and innovation programme under grant agreement No. $680843$.

\bibliographystyle{splncs04}
\bibliography{mybibfile}

\end{document}